# Motion of Dust Ejected from the Surface of Asteroid (101955) Bennu

Yu Jiang[1,2,3], Jürgen Schmidt[3]

1. State Key Laboratory of Astronautic Dynamics, Xi'an Satellite Control Center, Xi'an, China
2. School of Aerospace Engineering, Tsinghua University, Beijing 100084, China
3. Astronomy Research Unit, University of Oulu, Oulu, Finland

Email: jiangyu_xian_china@163.com

Abstract:
From Jan. 6, 2019 to Feb. 18, 2019, OSIRIS-REx observed asteroid (101955) Bennu ejecting 11 plumes of dust, of which part is escaping and another part is re-captured by the asteroid. The relative magnitudes of the typical forces acting on the emitted dust are quite different from the environments of the planets and other minor planets in the solar system. Here we show that ejected dust grains from the surface of Bennu can be caught in the gravitational field of Bennu. To this end, we calculated numerically the trajectories of dust grains of various sizes, from the 0.1 μm to the ten millimeter range. The shape and the fate of an emitted cloud of particles depend on the size of the grains: smaller grains form a more narrowly confined dust trail while trails formed by larger grains disperse more rapidly. Four different fates are possible for ejected dust. All grains with radius less than 1.0μm, directly re-impact on Bennu or they escape directly. In contrast, a fraction of grains with a radius larger than 10.0 μm will impact or escape only after performing a number of non-Keplerian revolutions around Bennu. Our findings show how dust grains may populate the vicinity of Bennu and other active asteroids and that they can reach interplanetary space and other celestial bodies, implying that organic matter can be transported from carbonaceous asteroids to other celestial bodies, including Earth.

# 1. Introduction

The properties of asteroid (101955) Bennu are consistent with a rubble-pile structure forming a microgravity aggregate[1]. Bennu is a carbonaceous asteroid belonging to the Apollo group with a semi-major axis of 1.12639AU [2,3]. The mean diameter of Bennu is 490.06 ±0.16 m [1]. The observation of the Origins, Spectral Interpretation, Resource Identification, and Security-Regolith Explorer (OSIRIS-REx) mission revealed that asteroids like Bennu may act as a class of bodies that have brought volatiles and organic compounds to Earth[2]. The OSIRIS-REx mission aims to return at least 60g of surface material back to Earth to study the role asteroids may have played for the origin of life on Earth, the results of which may help to distinguish between competing models for formation and evolution of the Solar System[4]. The dust grains are ejected from the surface at speeds ranging from a few centimeters per second to 3 meters per second[5,6]. Some of the particles escapes the gravitational field of Bennu while others remain bound[6]. The boulders on the surface of Bennu are observed to be covered by dark material which is interpreted as low-albedo dust[7].

The parameters of Bennu we use here are as follows: the rotation period is 4.296057h [1], the rotation axis points towards (85.65°, –60.17°) in J2000 equatorial coordinates[1,8], and the bulk density is $1,190 \pm 13 \, kg \cdot m^{-3}$ [1,8]. The Bennu shape model is the 75-centimeter shape model (MAR. 2019) with 196608 faces, 294912 edges, and 99846 vertex[9]. The mass concentration (or mascon) of a celestial body mean a region of crust which contains a positive gravitational anomaly. We assume Bennu is a homogeneous body. Other alternatives may consider the gravitational acceleration for the homogeneous body and the body with a masscon. However, we do not know the internal structure of Bennu. Using the assumption of homogeneous body with bulk density to simulation the gravitational acceleration of Bennu has enough accuracy.

## 2. Methods

We assume a fictitious spacecraft's location is in the dust trail, the fictitious spacecraft has a dust collector. The dust collector can detect dust grains larger than a fixed size $s_f$. The flux of dust grains at the dust collector is

$$N_c = \frac{S_c}{\left(\frac{b}{d_e/2}\right) \times 4\pi \left(\frac{d_m}{2}\right)^2} N_d = \frac{S_c d_e}{2\pi b d_m^2} N_d, \tag{1}$$

Here $b$ is the distance between the spacecraft and the center of Bennu; $d_e$ represents the size of Bennu in the equator; $d_m$ represents the mean size of Bennu; $S_c$ represents the cross-sectional area of the dust collector; $N_d$ represents dust grains larger than the fixed size $s_f$ which escape Bennu per second; $N_c$ represents the number of detections per second detected by the dust collector with the size larger than $s_f$. For Bennu, $d_e$ =563.78 m and $d_m$ =529.14 m are calculated by the polyhedron model.

We calculate the trajectories by numerically integrating the equations of motion for dust grains. The dust grains are generated on the surface of Bennu. The equations of motion for a dust grain can be written as

$$\ddot{\mathbf{r}} + 2\boldsymbol{\omega} \times \dot{\mathbf{r}} + \boldsymbol{\omega} \times (\boldsymbol{\omega} \times \mathbf{r}) + \dot{\boldsymbol{\omega}} \times \mathbf{r} = \mathbf{f}_A + \mathbf{f}_S + \mathbf{f}_P + \mathbf{f}_L + \mathbf{f}_{IM} + \mathbf{f}_{SR} + \mathbf{f}_{PR} + \mathbf{f}_{PD}, \tag{2}$$

where $\mathbf{r}$ is the radius vector of the dust relative to the mass centre of the asteroid; the first and second time derivatives of $\mathbf{r}$ are expressed relative to the asteroid's body-fixed frame; $\boldsymbol{\omega}$ and $\dot{\boldsymbol{\omega}}$ represent the rotational angular velocity vector and angular acceleration velocity vector, respectively; $\mathbf{f}_A$ represents the acceleration term caused by the asteroid's gravity; $\mathbf{f}_L$ represents the Lorentz acceleration caused by the magnetic field of Bennu; $\mathbf{f}_{IM}$ represents the Lorentz acceleration caused by the interplanetary magnetic field, $\mathbf{f}_{SR}$ represents the acceleration term caused by the solar radiation pressure; $\mathbf{f}_{PR}$ represents the acceleration term caused by the Poynting-Robertson drag; $\mathbf{f}_{PD}$ represents the acceleration term caused by the plasma drag; $\mathbf{f}_S$ represents the gravitational acceleration term caused by the Sun, and $\mathbf{f}_P$ represents the gravitational acceleration term caused by planets. The magnitude of the different accelerations[10] are different based on the parameters of the dust grains, such as the size, area-mass ratio, relative distance to the minor celestial bodies, etc.

The normal escape velocity is calculated by the following equation:

$$v_{esc}(\mathbf{r}) = -\mathbf{n}\cdot(\boldsymbol{\omega}\times\mathbf{r}) + \sqrt{[\mathbf{n}\cdot(\boldsymbol{\omega}\times\mathbf{r})]^2 + 2U(\mathbf{r}) - (\boldsymbol{\omega}\times\mathbf{r})^2}, \qquad (3)$$

Here $\mathbf{n}$ is the unit vector along the normal direction.

The gravitational acceleration can be calculated by the polyhedral model[11]:

$$\mathbf{f}_A = -G\sigma\cdot\mathbf{d}\sum_{e\in edges}\mathbf{E}_e\bullet\mathbf{r}_e\cdot L_e + G\sigma\cdot\mathbf{d}\sum_{f\in faces}\mathbf{F}_f\bullet\mathbf{r}_f\cdot\omega_f, \qquad (4)$$

where $G=6.67\times10^{-11}$ m$^3$kg$^{-1}$s$^{-2}$ is the gravitational constant, $\sigma$ is the bulk density of the asteroid; $\mathbf{d}$ represents the unit vector pointing from the dust grain to the mass centre of the asteroid; $\mathbf{E}_e$ is calculated by a appointed edge with two face- and edge-normal vectors; $\mathbf{F}_f$ is the face dyad calculated by $\mathbf{F}_f = \hat{\mathbf{n}}_f\hat{\mathbf{n}}_f$; $\hat{\mathbf{n}}_f$ is the face normal vector pointing to the outside of the surface. Body-fixed vectors $\mathbf{r}_e$ and $\mathbf{r}_f$ point from field points to any points on edge $e$ and face $f$, respectively. $L_e = \ln\dfrac{a+b+e}{a+b-e}$, $a$ and $b$ are distances from field point to the two ends of the edge, $e$ is the edge length; $\omega_f$ is the signed solid angle. The symbols • and · represent the inner and scalar product, respectively.

The Lorentz acceleration is

$$\mathbf{f}_L = \frac{q}{m}\dot{\mathbf{r}} \times \mathbf{B}_a, \qquad (5)$$

$$\mathbf{f}_{IM} = \frac{q}{m}\dot{\mathbf{r}} \times \mathbf{B}_{IM}, \qquad (6)$$

where $q$ is the charge of the dust; $m$ is the mass of the dust; $\mathbf{B}_a$ is the magnetic flux density of the asteroid; $\mathbf{B}_{IM}$ is the magnetic flux density of the solar wind.

The acceleration term caused by the solar radiation pressure[12] is

$$\mathbf{f}_{SR} = \left(\frac{S_0 A}{mc}\right)\left(\frac{r_{s0}}{r_{sd}}\right)^2 Q_{pr}\mathbf{r}_{SR}, \qquad (7)$$

where $S_0 = 1.3608 \times 10^6$ ergs·cm$^{-2}$·s$^{-1}$ is the solar constant; $A$ is the cross-sectional area of the dust; $c$ is the velocity of light; $r_{s0} = 1\mathrm{AU}$; $r_{sd}$ is the distance from the Sun to the dust; $Q_{pr} = Q_{ext} - Q_{sca}\cos\theta_s$ is the radiation pressure coefficient (see Fig. 1) and can be calculated using Mie theory[13-15]; $Q_{ext}$ and $Q_{sca}$ are the factors for extinction and scattering, respectively; $\cos\theta_s$ is the asymmetry factor related to the scattered radiation; $\mathbf{r}_{SR}$ is the photon beam's direction. Bennu's shadow is calculated by the following method: Let r be the norm of $\mathbf{r}$, i.e., the distance from the mass center of the asteroid to the dust grain, let $\mathbf{r}_0$ on the line from the Sun to the dust grain, and $\mathbf{r}_0 = \mathbf{r} - 2r\frac{\mathbf{r}_{SR}}{|\mathbf{r}_{SR}|}$. Now we separate the line segment from $\mathbf{r}_0$ to $\mathbf{r}$, if there exist a point $\mathbf{r}_j$ located in the polyhedral model, then the dust grain is in Bennu's shadow.

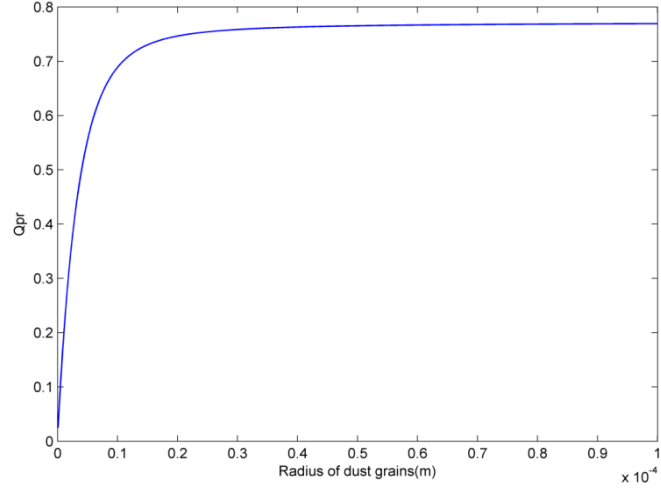

**Fig. 1 | The value for $Q_{pr}$.**

The acceleration term caused by the Poynting-Robertson drag[12] is

$$\mathbf{f}_{PR} = -\left(\frac{S_0 A}{mc^2}\right)\left(\frac{r_{s0}}{r_{sd}}\right)^2 Q_{pr} \dot{\mathbf{r}}_{sd}, \tag{8}$$

where $\mathbf{r}_{sd}$ and $\dot{\mathbf{r}}_{sd}$ are the position and velocity vectors from the Sun to the dust, respectively.

The acceleration term caused by the plasma drag[16] is

$$\mathbf{f}_{PD} = -\pi n_H \frac{m_H}{m} s^2 v_{pd} \mathbf{v}_{pd}, \tag{9}$$

where $n_H$ is the number density of heavy ions (O+, OH+, H2O+, H3O+, and N+); $m_H$ is the mass of heavy ions; $s$ is the diameter of the dust; $\mathbf{v}_{pd}$ is the velocity vector of the dust relative to the plasma; $v_{pd} = |\mathbf{v}_{pd}|$. The molar mass of heavy ions is assumed to be 17, thus the average mass of a heavy ion is set to be $m_H = 17/(6.02\times10^{23}) = 2.824\times10^{-23}$ kg. The number density of heavy ions is $n_H = 5\times10^6 \mathrm{m}^{-3}$ [17].

The charge of a dust grain is calculated by

$$q = 4\pi\varepsilon_0 U r_d, \tag{10}$$

where $\varepsilon_0 = 8.854187817 \times 10^{12} F \cdot m^{-1}$ is the permittivity; $U = +5V$ is the surface potential of the dust in the interplanetary plasma environment; $r_d$ is the radius of a spherical dust grain.

The magnetic flux density[16-20] of the asteroid is calculated by a simple dipole model:

$$B_a = \mu_0 \frac{M}{3} \left( \frac{r_a}{r} \right)^3, \tag{11}$$

where $B_a$ is the norm of the magnetic flux density $\mathbf{B}_a$; $M$ is the magnetization of Bennu; $r_a = \frac{d_m}{2}$ is the mean radius of Bennu; $r$ is the norm of $\mathbf{r}$; the magnetic pole is assumed to be parallel with the rotational axis of Bennu.

We use the magnetic field of asteroid Braille as an analogue for Bennu. Although Braille is a Q type, but Bennu is a C type asteroid. It maybe hard to measure magnetic field of minor celestial bodies. For instance, ROSETTA spacecraft performed a flyby of asteroid (21) Lutetia with the relative distance 3120 km, the magnetometers did not detect any conclusive signature of Lutetia[21]. The magnetization of Bennu is estimated by

$$M = M_{Br} \left( \frac{r_a}{r_{Br}} \right)^2, \tag{12}$$

where $M_{Br} = 2.1 \times 10^{11} A \cdot m^2$ is the maximum possible dipole moment for asteroid (9969) Braille; $r_{Br} = 0.78 km$ is the mean radius of Braille[18]. The effect of magnetic field from Bennu to dust grains is too small and is negligible.

The magnetic flux density[22] of the solar wind can be expressed in heliographic coordinates $\left( B_r, B_\phi, B_\theta \right)$ as

$$B_r = B_{r,0} \left(\frac{r_0}{r_{hg}}\right)^2$$

$$B_\phi = B_{\phi,0} \frac{r_0}{r_{hg}} \cos \beta_{hg}, \quad (13)$$

$$B_\theta = 0$$

where $B_{r,0} = B_{\phi,0} = 3\text{nT}$; $r_0 = 1\text{AU}$; $r_{hg}$ is the distance from the Sun to the dust grain; $\beta_{hg}$ is the latitude of the grain in heliographic coordinates.

The distribution density function of the initial velocity for dust ejected from the surface of the asteroid [23] is calculated by

$$f(\hat{v}) = \frac{\delta \hat{v}}{(1-\hat{v}^2)^2} e^{-\frac{\beta \hat{v}}{1-\hat{v}^2}}, \quad (14)$$

where $\hat{v} = v_{ejc}/v_{max}$, $v_{ejc}$ is the ejection velocity of a dust grain, $v_{max}$ is the maximum ejection velocity; $\beta = 8.69, \delta = 7.2 \times 10^{-3} \text{s} \cdot \text{m}^{-1}$. For ejected dust grains from the surface with the speeds interval[6] from a few centimeters per second to 3 meters per second, $v_{max} = 3\text{m} \cdot \text{s}^{-1}$. We neglect the ejection velocity smaller than $v_{min} = 0.02\text{m} \cdot \text{s}^{-1}$ and re-orthogonalize the distribution density function.

The mass distribution of the ejecta[23] is calculated by

$$N^+(>M_d) = \frac{1-\alpha}{\alpha} \frac{M^+}{M_{max}} \left(\frac{M_{max}}{M_d}\right)^\alpha, \quad (15)$$

where $M^+ = 0.15\text{kg} \cdot \text{s}^{-1}$ is the mass production rate for Bennu; $\alpha = 0.83$; $M_d$ is the mass of the dust grain, and $N^+(>M_d)$ is the number of grains with masses larger than $M_d$ ejected from the surface of Bennu per second.

The Yarkovsky effect[24] is a force caused by theanisotropic emission of thermal photons acting on asteroids, asteroids absorb energy from thermal photons and emit as heat, this leads to acceleration acting on asteroids. The acceleration caused by Yarkovsky effect is much smaller than gravitational accelerations from Sun and planets, but it can leads to large orbital changes of asteroids for millions to billions of years. For the physical phenomenon of the theanisotropic emission of thermal photons acting on asteroids, when it create a force, it is the Yarkovsky effect; when it create a thermal torque, it is the Yarkovsky-O'Keefe-Radzievskii-Paddack (YORP) effect acting on asteroids[25]. The YORP effect leads to the variety of the direction and value of the asteroid's rotation vecoclty. Here we simulate the emmision of dust grains, the timescales is smaller than 2 days, and much smaller than millions to billions of years. Thus we neglect the Yarkovsky and YORP effects.

## 3. Results

The maximum and minimum normal escape velocities on the surface of Bennu are 0.178 m·s$^{-1}$ and 0.00534 m·s$^{-1}$, respectively (see Fig.2). The escape velocity is along the normal direction of the surface. The meteoroid impact flux of Bennu at different points in its orbit around the Sun are different [26], the ejection of particles has a long-term effect on Bennu's orbit, attitude, and regolith distribution, detailed contents of the activity research of Bennu can be seen in Hergenrother et al. [26]. Here we considered the rotation of the body and the shape model[1,9]. Table 1 presents assumed physical parameters for simulation. Assuming the grain production[5] rate 150g·s$^{-1}$, this value is an upper bound averaged from a single 34-minute measurement period for comet 67 P at perihelion. Then the numbers of grains with the size larger than 1 cm , 1mm, 100μm, 10μm, 1μm, and 0.1μm are 5,

1700, $5.2\times10^5$, $1.6\times10^8$, $5.0\times10^{10}$, and $1.5\times10^{13}$ per second (see Eq. (15)). The minimum and maximum ejection velocities[6] are modeled as 0.02 m·s$^{-1}$ and 3.0 m·s$^{-1}$, respectively, following a velocity distribution inferred for ejecta from the Moon[27]. The numbers of ejected dust grains per second with their eventual fate (impact/escape) are presented in Table 2. The simulation code had been developed by us using Fortran. Among 1700 grains with the size larger than 1mm, the numbers of grains that impact on the surface and escape are 198 and 1502, respectively. Figure 3 shows the dust flux vs. the distance from the mass center of Bennu, the grains are langer than 0.1μm. The value of the minimum detectable size and the cross-sectional area of the dust collector of a fictitious spacecraft are presented in Table 1. The fictitious spacecraft's location is in the dust trail and the distance between the spacecraft and the center of Bennu is 1.0 km, then the dust collector would detect approximately $5\times10^4$ detections per second.

Table 1| Assumed Physical Parameters for Simulation.

| | |
|---|---|
| grain productionrate[5] | 150g·s$^{-1}$ |
| Minimum ejection velocity[6] | 0.02 m·s$^{-1}$ |
| maximum ejection velocity[6] | 3.0 m·s$^{-1}$ |
| minimum detectable size of the dust collector of a fictitious spacecraft | 0.1 μm |
| cross-sectional area of the dust collector of a fictitious spacecraft | 0.01m$^2$ |

Table 2| Numbers of Ejected Dust Grains per Second with Their Eventual Fate (Impact/Escape), the results come from numerical integration of trajectories of dust grains.

| Radius of Dust Grains | Total Number | Impact | Escape |
|---|---|---|---|
| ⩾0.1μm | $1.5\times10^{13}$ | $3.3\times10^{12}$ (2.2%) | $1.17\times10^{13}$ (97.8%) |
| ⩾1.0 μm | $5.0\times10^{10}$ | $5.8\times10^9$ (11.6%) | $4.42\times10^{10}$ (88.4%) |
| ⩾10.0 μm | $1.6\times10^8$ | $1.57\times10^7$ (9.81%) | $1.443\times10^8$ (90.19%) |
| ⩾100.0 μm | $5.2\times10^5$ | $6\times10^4$ (11.54%) | $4.6\times10^5$ (88.46%) |
| ⩾1.0 mm | 1700 | 198 (11.65%) | 1502(88.35%) |

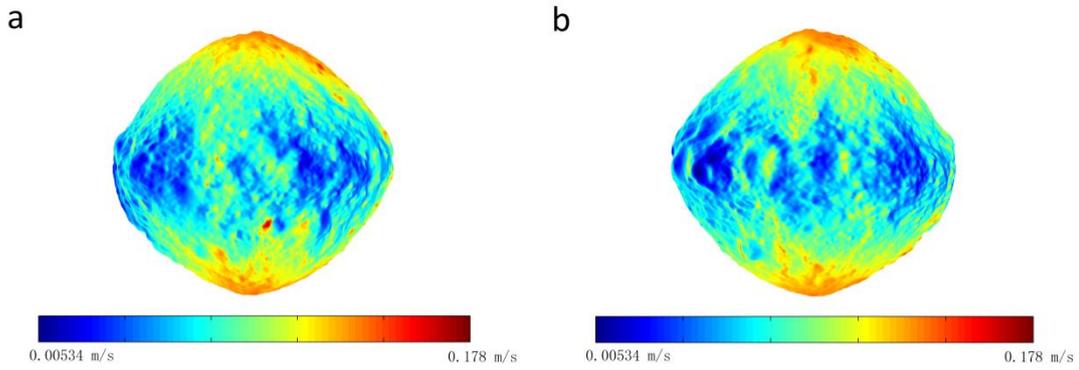

**Fig. 2| Escape velocity on the surface of Bennu**. a: viewed from +x axis; b: viewed from –x axis. The axes are defined by: x is the axis of minimum inertia, z is the axis of maximum inertia, y and z, x are right-handed.

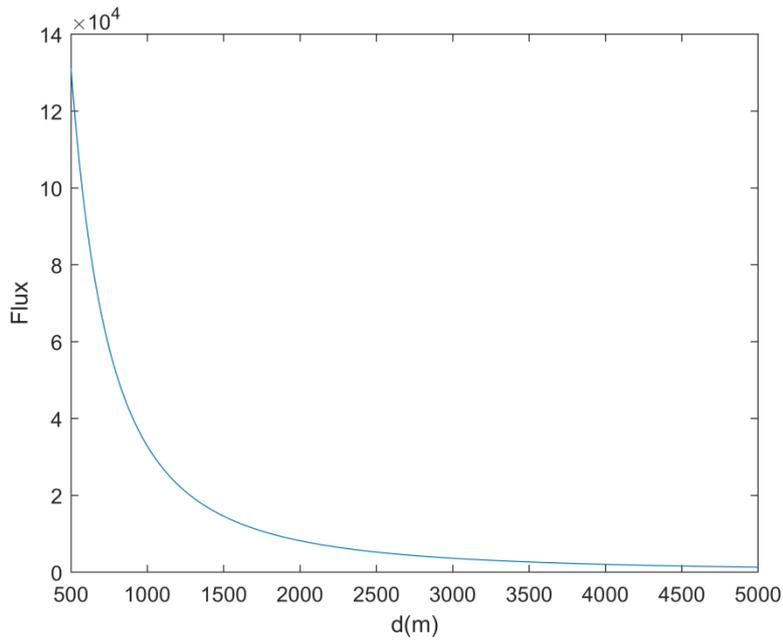

**Fig. 3| The Dust Flux vs. the Distance from the Mass Center of Bennu.** the grains are langer than 0.1μm.

For asteroids, the micrometeoroids are the major causes of ejected particles, including helion and anti-helion sources[28], north and south apex sources[29], and north and south toroidal sources[30]. Dust grains can be generated by rotational surface shedding, breakup, etc. Here we only analyze the dust grains generated by the impact ejection. Figure 4 presents trajectories of 2000 ejected dust grains from the surface of asteroid (101955) Bennu. The ejection velocities are also simulated from 0.02 m·s$^{-1}$ to 3.0 m·s$^{-1}$ with a velocity distribution[6,27]. The maximum and minimum sizes of dust grains we simulate are 10.0 mm and 0.1μm, respectively, following a power law size distribution[27,31-35]. The ejection angles are evenly distributed in a range of [30.0 75.0] deg. A 0 deg ejection is a nornal ejection. In Figure 4, the dust grains are generated by the distribution density function of the velocity and the mass distribution of the ejecta. It is assumed that micrometeoroids impact to the asteroid surface from normal direction. Thus we exclude normal ejections. Among the ejection generation, all the local surface region may generate dust grains, thus we use a uniform distribution to generate the initial positions of dust grains on the surface. This implies different local surface regions have the same probability to generate dust grains. The acceleration of the grains caused by a possible asteroidal magnetic field of reasonable magnitude is much smaller than the effect of solar radiation, and thus the effect is negligible (although we include the respective acceleration methods). In this case, all the dust grains will escape or impact onto Bennu directly, none of them can be caught in the gravitational field of Bennu. 477 dust grains will impact onto Bennu and other 1523 dust grains will escape directly. Among these 2000 grains, no grain are seen orbiting the asteroid because using the power law size distribution, most of the grains have small size, and the effect of the Solar radiation is large. The motion direction for dust grains is opposite on the Sun direction.

The velocity and size of ejecta dust have distributions[27,31-35], which means the numbers of different sizes of dust grains are different in the above simulation. To investigate the motion of different sizes of dust grains, we fix the grain size and consider the velocity distributions of dust. Bennu is a top-shaped asteroid, thus the gravitational acceleration on the surface at the same latitude are almost equal. We now choose one surface point to calculate the motion of grains with different size. The initial position is generated using a uniform distribution. For the size and mass of dust grain, we simulate dust grains with the radius of 0.1μm, 1.0 μm, 10.0 μm, 100.0 μm, 1.0 mm, and 10.0 mm. For each size, we calculate 2500 dust grains. The initial velocity is set to be 0.02m·s$^{-1}$ to 3.0 m·s$^{-1}$, and satisfies the ejection velocity distribution, although the observed velocity is only for grains in the cm to dm range because smaller grains can not be seen[28]. We calculated the force acceleration acting on the dust including the gravitational acceleration generated by the 75-centimeter shape model with constant-density, the acceleration term caused by the solar radiation pressure, the acceleration term caused by the Poynting-Robertson drag, gravitational acceleration term caused by the Sun and eight planets. The trajectories of dust grains with different radius are calculated (see Fig.5). In Figure 5, the dust grains are generated by the distribution density function of the velocity and the sizes set to be several different sizes, from submicron to the ten millimeter range. The different final state of these dust grains with fixed sizes are presented in Table 3. For dust grains with the radius of 0.1μm, 1.0 μm, 10.0 μm, 100.0 μm, 1.0 mm, and 10.0 mm, the longest temporal simulations to obtain results showed in Table 3 and

Fig.5 are $9.5\times10^2$, $3.3\times10^3$, $7.38\times10^4$, $8.8\times10^4$, $1.0315\times10^5$, and $1.273\times10^5$, respectively. Most of the dust grains will escape directly after they leave the surface of Bennu. For dust grains with the radius of 0.1μm, 75.44% of them will escape directly, and other 24.56% will impact onto the surface of Bennu directly. For dust grains with the radius of 1μm, 88.48% of them will escape directly, and other 11.52% will impact onto the surface of Bennu directly. For the radius of dust grains smaller than 1μm, all the dust will escape directly or impact onto Bennu directly, none of them can be caught in the gravitational field of Bennu. For the radius of 10μm, these two values are 90.24% and 9.2%, respectively; In addition, among 2500 grains, 14 dust grains will get caught in the gravitational field of Bennu, after several circles of orbits, eight of them will impact onto Bennu, and other six will escape. For dust grains with the radius of 100.0μm, in our 2500 grains, 21 dust grains will get caught in the gravitational field of Bennu, after several circles of orbits, 15 of them will impact onto Bennu, and other six will escape; In addition, 89% of dust will escape directly, and 10.16% of dust will impact onto Bennu directly. For dust grains with the radius of 1.0mm, 34 grains among 2500 grains will get caught in the gravitational field of Bennu, 24 of them will finally impact onto Bennu after several circles of orbits, the other ten will escape after several circles of orbits; 88.68% of them will escape directly, and 9.96% of them will impact directly. For grains with the radius of 10.0mm, which should not be considered to be dust, 47 grains among 2500 grains will get caught in the gravitational field of Bennu, 36 of which will final escape and the other eleven will finally impact onto Bennu; In addition, 88.08% of the total population will escape directly, and 10.04% of the grains will impact onto Bennu directly. Our calculation shows the major forces acting on the dust grains are gravitational acceleration from Bennu and the solar radiation. These particles from the surface can be captured into orbit at least for several periods because of the combined effect of the irregular shape as well as the gravitational acceleration from Bennu and the solar radiation. We find orbiting particles in this simulation in Fig.5, but not in the simulation in Fig.4, because in Fig.4, we use a power law size distribution to generate the dust grains, large sizes of dust grains are less in the simulation in Fig.4. The grains with the size larger are more likely to be captured into orbit. The norm and direction of the initial ejection velocity also have influence on the motion state of the grains. If the norm of the initial ejection velocity is too large or the direction is perpendicular to the local surface of Bennu, the grains will be hard to get captured into orbit. The grains that got captured into orbit may form asteroid ring if the generation of dust grains on the surface of asteroid is enough.

Table 3| Dust Grains with Different Final States, the results come from numerical integration of trajectories of dust grains.

| Radius of Dust Grains | Directly Impact(Less than one circle) | Impact After Several Circles of Orbits | Escape After Several Circles of Orbits | Directly Escape |
|---|---|---|---|---|
| 0.1μm | 614 (24.56%) | 0 | 0 | 1886 (75.44%) |
| 1.0 μm | 288 (11.52%) | 0 | 0 | 2212 (88.48%) |
| 10.0 μm | 230 (9.2%) | 8 | 6 | 2256 (90.24%) |
| 100.0 μm | 254 (10.16%) | 15 | 6 | 2225 (89 %) |
| 1.0 mm | 249 (9.96%) | 24 | 10 | 2217 (88.68%) |
| 10.0 mm | 251 (10.04%) | 36 | 11 | 2202 (88.08%) |

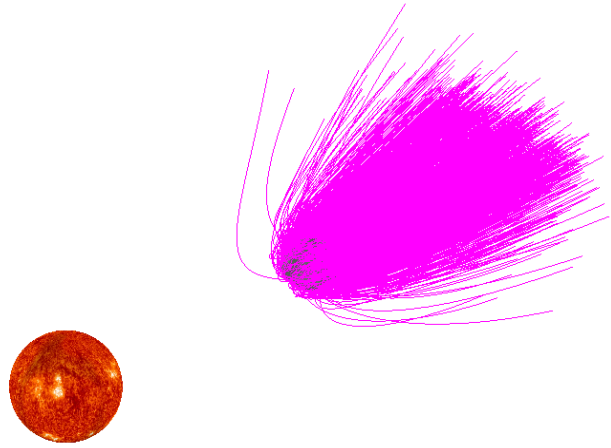

**Fig. 4 | Trajectories of dust around Bennu.** The trajectories are plotted relative to the inertial space. Dust grains are ejected from the asteroid surface. 2000 dust grains are generated. To see the influence of the solar radiation on the motion of dust grains, the Sun is plotted in the solar direction.

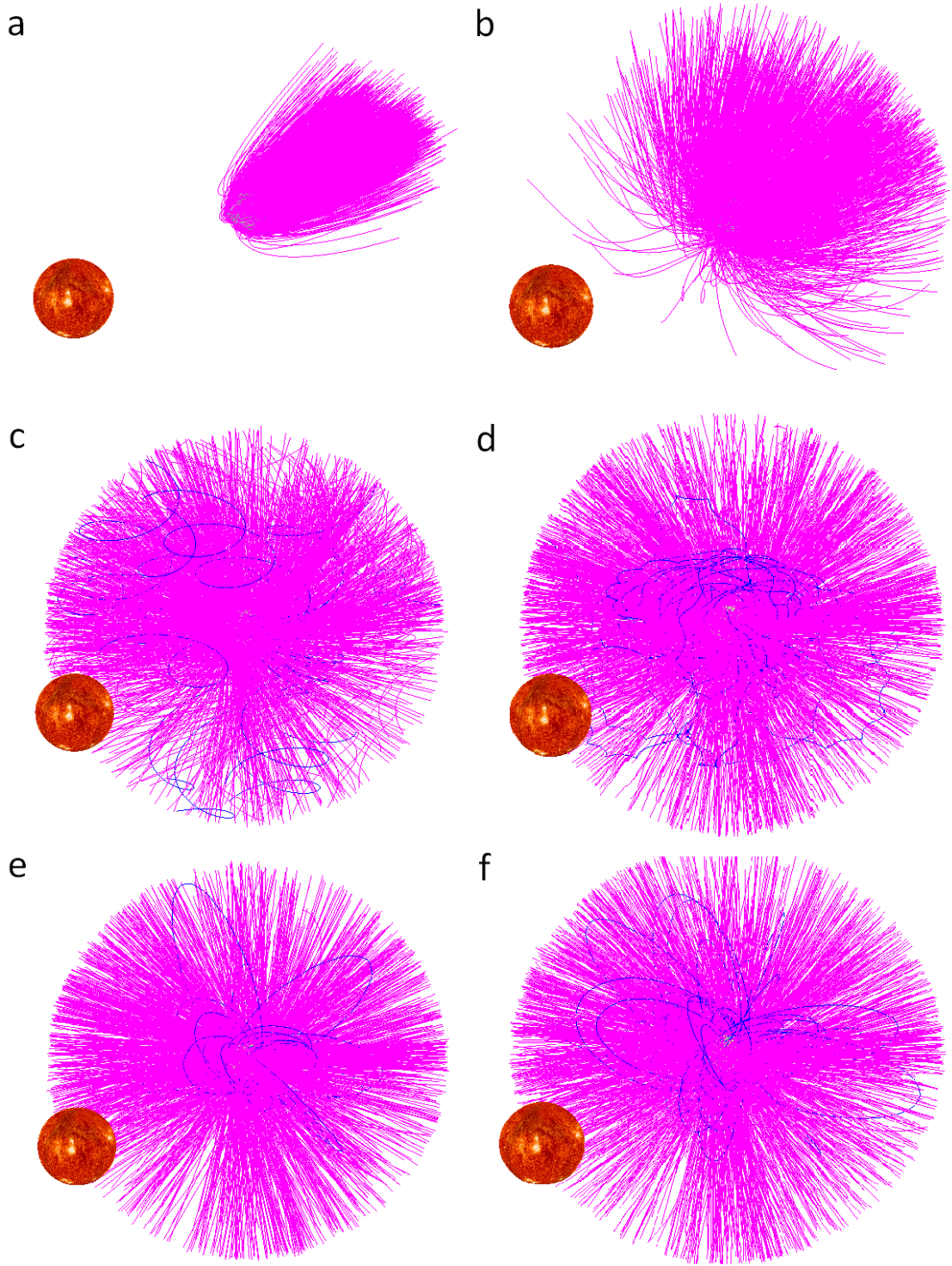

**Fig. 5 | Trajectories of dust grains with different radii.** The trajectories are plotted relative to the inertial space. Dust grains are ejected from the asteroid surface. For each size of dust grains, 300 dust grains are generated. Subfigures and radii, **a**: 0.1μm; **b**: 1.0 μm; **c**: 10.0 μm; **d**: 100.0 μm; **e**: 1.0 mm; **f**: 10.0 mm.

# 4. Conclusion

Each impact of micrometeoroids generates lots of dust grains. The study of dust emmision can help us to understand the material migration from asteroids to Interplanetary space, the evolution and reconstructed shape of asteroid shape, the formation of asteroid trail, etc. The dust trails are generated by all these dust grains generated by these impacts. The direction and shapes of the dust trails depend on the effect of the solar radiation, the gravity of the asteroid and the Sun, and the Lorentz force, it also related to the charge, size, and density of the dust grains. For an asteroid, the direction of dust trails maybe in the anti-solar direction or along its orbit; the direction depends on the velocity of the dust and the orbital velocity of the asteroid. For Bennu, among escaped dust, large dust grains' directions are mainly along its orbit, and small dust grains' directions are mainly in the anti-solar direction. When we used higher ejection velocities for smaller particles, more time is needed for particles to change the directions of orbits.

If the size of dust grains becomes large, the field angles of dust trails formed by these dust grains also becomes large. In other words, larger grains form a more widely scattered dust trail. This implies that the dust trail formed by grains with small size looks thin while the dust trail formed by grains with large size looks thick. The motion of ejected dust grains forms dust trails which have different sizes of grains in the trails. When the charge-to-mass ratios for different dust grains are the same, the trajectories of the dust grains with large size are different from the trajectories of the dust grains with small size. In the interior of the dust trails, the sizes of dust grains are smaller than the sizes of dust grains near the external regions.

Referring to the shapes of trajectories of dust grains which can get caught in the gravitational field of Bennu, one can find in Fig. 5 that the dust grains with the radius of 10.0μm and 100.0μm which were caught around Bennu have S-shaped trajectories. For the dust grains with the radius of 1.0 mm, the trajectories of dust grains which can get caught in the gravitational field of Bennu look a little curved; however, the curvature is smaller than that for grains with the radius of 100.0μm. When the radius of the grains equal to 10.0 mm, the trajectories of dust grains which can get caught in the gravitational field of Bennu do not have S-shaped trajectories like the trajectories of grains with the radius of 100.0μm.

The simulation indicated that when the size of ejected dust grains larger than 10μm may be caught in the gravitational field of Bennu. Although most of them will directly escape or impact to Bennu. Our finding shows that the shape and the fate of an emitted cloud of particles depend on the size of the grains, i.e., smaller grains form a more narrowly confined dust trail. If the size of dust grains become large, dust trail will disperse more rapidly. This research is useful for understanding the generation of dust grains around Bennu and other active asteroids, the shape change of asteroids, as well as the the material transfer from asteroids to interplanetary space and other celestial bodies, including Earth.